\documentclass[11pt]{article}

\usepackage{graphicx,amssymb}

\long\def\inst#1{\par\nobreak\kern 4pt\nobreak
    {\it #1}\par\vskip 10pt plus 3pt minus 3pt}

\def\missPT{{P\mkern-11mu\slash}_T}

\begin{document}

\begin{center}
{\Large \bf \boldmath
Signatures of black holes at the LHC\footnote{Work supported in part by the U.S. Department of Energy contract DE-FG05-91ER40622.}} \\
\vspace*{1.0cm} 
{M.\ Cavagli\`a${}^a$, R.\ Godang${}^{a,b}$, L.\ M.\ Cremaldi${}^a$ and D.\ J.\ Summers${}^a$\\
\vspace*{0.3cm}   
{\em ${}^a$ Department of Physics and Astronomy,
University of Mississippi\\
University, MS 38677-1848, USA\\~\\
${}^b$ 
Department of Physics,
University of South Alabama,
Mobile, AL 36688, USA}}
\end{center}
\bigskip

\begin{center}
\bf Abstract
\end{center}
\vspace*{1.0cm}
{Signatures of black hole events at CERN's Large Hadron Collider are discussed. Event simulations
are carried out with the Fortran Monte Carlo generator CATFISH. Inelasticity effects, exact
field emissivities, color and charge conservation, corrections to semiclassical black hole
evaporation, gravitational energy loss at formation and possibility of a black hole remnant are
included in the analysis.}

\section{Introduction}
If the fundamental scale of gravity is of the order of few TeVs \cite{Arkani-Hamed:1998rs},
proton-proton collisions at CERN's LHC could lead to the formation of mini Black Holes (BHs)
\cite{Banks:1999gd} and branes \cite{Ahn:2002mj} (For reviews and further references, see Refs.\
\cite{Cavaglia:2002si, Cardoso:2005jq}). The cross section for creation of a BH or brane with radius
$R$ is expected to be approximately equal to the geometrical Black Disk (BD) cross section
$\sigma_{BD}(s,n)=\pi R^2(s,n)$, where $\sqrt{s}$ is the Center of Mass (CM) energy of the colliding
quanta and $n$ is the number of extra dimensions. The semiclassical Hawking effect
\cite{Hawking:1974sw} provides a decay mechanism for BHs which makes them visible to a detector. The
spectrum of massive excitations in string theories suggests that branes may also decay thermally
\cite{Amati:1999fv}. Under the most favorable circumstances, the BH event rate at the LHC should be
comparable to the $t\bar t$ event rate. 

Until now, numerical studies of observational signatures have implemented the semiclassical picture
outlined above. However, recent results have significantly modified our understanding of BH
formation and evolution. It is thus timely and worthwile to examine the observational signatures of
BH events beyond the simple semiclassical picture. To this purpose, we have analyzed BH events at
the LHC with the Fortran Monte Carlo (MC) generator CATFISH, which implements many of the accepted
theoretical results in the literature \cite{Cavaglia:2006uk,Gingrich:2006gs} and allows the
comparison of different theoretical models of BH production and decay. MC generators with similar
characteristics of CATFISH have already been successfully utilized to simulate BH production in
ultrahigh-energy cosmic ray air showers \cite{Ahn:2005bi} and in lepton colliders
\cite{Godang:2004bf}. 
\section{A quick look at the physics of mini black holes}
Thorne's hoop conjecture \cite{hoop} states that an event horizon forms when a mass $M$ is compacted
into a region with circumference smaller than twice the Schwarzschild radius $R(M)$ in any
direction. At the LHC, this process can be achieved by scattering two partons with CM energy larger
than $M$ and impact parameter smaller than $R$. Analytic and numerical results show that the BH
event is inelastic due to emission of gravitational radiation \cite{Cardoso:2005jq}. If the
collision is elastic, the hoop conjecture implies that the parton cross section for BH production is
equal to the geometrical cross section $\sigma_{BD}$. Otherwise, the cross section is smaller and
depends on the impact parameter. The collisional energy loss depends on the impact parameter and
increases as the number of spacetime dimensions increases. Consensus is that the BH mass
monotonically decreases with the impact parameter from a maximum of about 60-70\% of the CM energy
for head-on collisions \cite{Yoshino:2002tx,Yoshino:2005hi,Vasilenko:2003ak}. However, other
independent estimates suggest that the gravitational energy loss could be smaller
\cite{Cardoso:2002pa,Berti:2003si}. Note that these treatments are rigorous only for BHs larger than
the Compton length of the colliding quanta \cite{Voloshin:2001vs}. Moreover, mass, spin, charge and
finite-size effects of the incoming partons are neglected. Size and spin effects are expected to be
mostly relevant around the Planck energy. Charge effects could dominate at higher energy. The
pointlike approximation fails for directions transversal to the motion \cite{Kohlprath:2002yh}. 

The total cross section for a super-Planckian BH event involving two nucleons
is obtained by integrating the parton cross section over the Parton
Distribution Functions (PDFs). If the BH mass depends on the impact parameter, the
generally accepted formula for the total cross section in a proton-proton collision is
\begin{equation}
\sigma_{pp \to BH}(s,n) = \sum_{ij}\int_0^1 2z dz
\int_{x_m}^1dx\,\int_x^1 {dx'\over
x'}\,f_i(x',Q)f_j(x/x',Q)\,F\,\sigma_{BD}(xs,n)\,,
\label{totcross}
\end{equation}
where $f_i(\cdot,Q)$ are the PDFs with four-momentum transfer squared $Q$
\cite{Brock:1993sz,pdg2006} and $z$ is the impact parameter normalized to its maximum value. The
cutoff at small $x$ is $x_m=M_{min}^2/(sy^2(z))$, where $y(z)$ and $M_{min}$ are the fraction of CM
energy trapped into the BH and the minimum-allowed mass of the gravitational object, respectively.
$F$ is a form factor. The total cross section for the BD model is obtained by setting $F=1$ and
$y^2(z)=1$. The momentum transfer is usually set to be $M_{BH}$ or the Schwarzschild radius inverse.
The lower cutoff on the fraction of the nucleon momentum carried by the partons is set by the
minimum-allowed formation mass of the gravitational object, $M_{min}$. This threshold is usually
considered to be roughly equal to the minimum mass for which the semiclassical description of the BH
is valid. However, this argument is based on Hawking's semiclassical theory and may not be valid at
energies equal to few times the Planck mass. For example, the existence of a minimum spacetime
length $l_{m}$ implies the lower bound on the BH mass \cite{Cavaglia:2003qk,Cavaglia:2004jw}:
\begin{equation}
M_{ml}= \frac{n+2}{8\Gamma\left(\frac{n+3}{2}\right)}
\left(\sqrt{\pi}\,l_{m}M_\star/2\right)^{n+1} \, M_{\star}\,,
\label{minmass}
\end{equation}
where $M_{\star}$ is the fundamental Planck mass. BHs with mass less than $M_{ml}$ do not exist, since
their horizon radius would fall below the minimum-allowed length.

After its formation, the mini BH is believed to radiate excess multipole moments (balding phase),
spin-down and then classically evaporate through the Hawking mechanism. At the end of the Hawking
evaporation, the BH may undergo a non-thermal decay in a number $n_p$ of hard quanta or leave a
remnant. Although some progress has been made, a complete quantitative description of the BH
evolution is not fully known. The better understood stage is the Hawking phase, for which
(classical) field emissivities have recently been calculated for {\it all} Standard Model (SM)
fields \cite{Cardoso:2005vb}. (For earlier works on spin-0, -1/2 and -1 fields see Refs.\
\cite{Kanti:2002nr}.) For the minimal $SU(3)\times SU(2)\times U(1)$ SM, most of the BH mass is
radiated as SM quanta on the brane, although the gravitational emission in the bulk cannot be
neglected for high $n$. It should be stressed, however, that the effect of rotation and quantum
corrections on BH emissivities is not clear. Onset of additional evaporation channels at
trans-Planckian energies could also lead to a larger emission of undetectable non-SM quanta during
the decay phase even in absence of rotation \cite{Cavaglia:2003hg,Chamblin:2004zg}. Quantum
gravitational effects and BH recoil \cite{Frolov:2002as} could also affect the emission of visible
quanta on the brane. Examples of quantum gravitational effects are quantum thermal fluctuations and
corrections to Hawking thermodynamics due to the existence of a minimum length
\cite{Cavaglia:2004jw}. In absence of a BH remnant, the final non-thermal decay is usually
described phenomenologically by setting a cutoff on the BH mass of the order of the Planck mass,
$Q_{min}\sim M_\star$, and democratically distributing the energy to the quanta. The existence of a
minimum length gives a natural means to set $Q_{min}$. In this case, the modified thermodynamical
quantities determine the endpoint of Hawking evaporation when the BH mass reaches $M_{ml}$. 
\section{The CATFISH generator}
In this section we review the main characteristics of the CATFISH generator. CATFISH includes three
models for BH formation and cross section: BD, Yoshino-Nambu (YN) graviton loss model
\cite{Yoshino:2002tx}, and Yoshino-Rychkov (YR) graviton loss improved model \cite{Yoshino:2005hi}.
Since the differences between the YN and YR models are not significant, only the latter has been
used in the analysis below. The distribution of the initial BH masses is sampled from the
differential cross section. CATFISH uses the cteq5m1 PDF distribution \cite{Brock:1993sz,
Lai:1999wy}. (The use of different PDF distributions does not significantly affect the total and
differential cross sections. For a detailed discussion on the uncertainties in the cross section
due to the PDFs, see Ref.\ \cite{Ahn:2003qn}.)  Following earlier studies
\cite{Anchordoqui:2001cg}, the momentum transfer is set to $Q =
\hbox{min}\,\{M_{BH}~\hbox{or}~R(M_{BH}), Q_{\rm max}\}$, where  $Q_{\rm max}$ is the maximum value
allowed by the PDFs. The part of CM energy of the pp collision which is not trapped or lost in
gravitational radiation forms the beam remnant, which is hadronized by PYTHIA
\cite{Sjostrand:2006za}. Energy losses in the balding and spin-down phases are assumed to be either
negligible or included in the energy loss during formation. 

Exact classical emissivities of non-rotating spherically-symmetric BHs are implemented in the
Hawking phase \cite{Cardoso:2005vb}. The particle content at trans-Planckian energy is assumed to
be the minimal $SU(3)\times SU(2)\times U(1)$ SM with three families and a single Higgs boson on a
thin brane. For black holes with mass $\sim$ few TeV the Hawking temperature is generally above
$100$ GeV. Therefore, all SM degrees of freedom are considered massless. Presence of a minimum
length may affect the evaporation phase and is implemented in CATFISH. The MC uses the
dimensionless parameter $\alpha=l_m M_\star/2$ to determine the minimum length
\cite{Cavaglia:2003qk,Cavaglia:2004jw}. If there is no minimum length, the MC evaporates the BH
according to the Hawking theory with varying temperature. Alternatively, the BH evolution proceeds
according to the modified thermodynamics of Ref.\ \cite{Cavaglia:2003qk,Cavaglia:2004jw}. The
evaporation ends with a stable BH remnant or an  explosive $n_p$-body decay when the BH reaches the
mass $Q_{min}$. Color charge is always conserved in the decay process. Conservation of EM charge
can be turned off to make the BH remnant electrically charged. Four-momentum is conserved at each
step in the evaporation process by taking into account the recoil of the BH on the brane due to the
emission of the Hawking quanta. The initial energy of the BH is distributed democratically among
all the Hawking quanta with a tolerance of $\pm 10$\%. Beam remnant, fragmentation, and initial-
and final-state radiation are dealt with PYTHIA. 
\section{Analysis of black hole events}
We focus on a purely statistical analysis of variables which allows an easy comparison with previous
results \cite{Stocker:2006we,Harris:2004xt,Alberghi:2006qr,Tanaka:2004xb,Lonnblad:2005ah,
Nayak:2006vf} which have been obtained with the TRUENOIR \cite{Dimopoulos:2001en} or CHARYBDIS
\cite{Harris:2003db} generators. A more refined analysis of other detector response-dependent
signatures such as back-to-back di-jet suppression, di-lepton events ($\mu^+\mu^-$, $\mu^+e^-$,
$\mu^+e^+$, \dots) will be presented in a future publication. 
\subsection{Visible and missing transverse momentum \label{vismiss}} 
\begin{figure}[ht]
\begin{center}
\centerline{\includegraphics[height=0.4\textheight,width=0.92\textwidth]{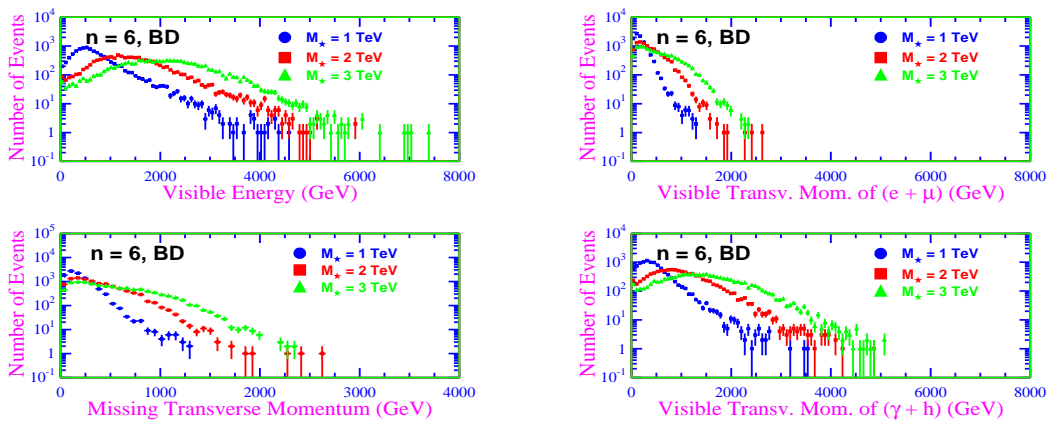}}
\vspace{-10truemm}
\caption[Visible energy, missing and visible transverse momentum for the BD model, $n=6$ and varying
$M_\star$.]{\label{figure:1} Visible energy, $\missPT$ and visible transverse momentum of leptons
and photons+jets (GeV) for the black disk model (BD) and fundamental Planck scale $M_\star=1,2,3$
TeV. The number of extra dimensions is $n=6$ and the final BH decay is in four hard quanta.}
\end{center}
\end{figure}

\begin{figure}[ht]
\begin{center}
\centerline{\includegraphics[height=0.4\textheight,width=0.92\textwidth]{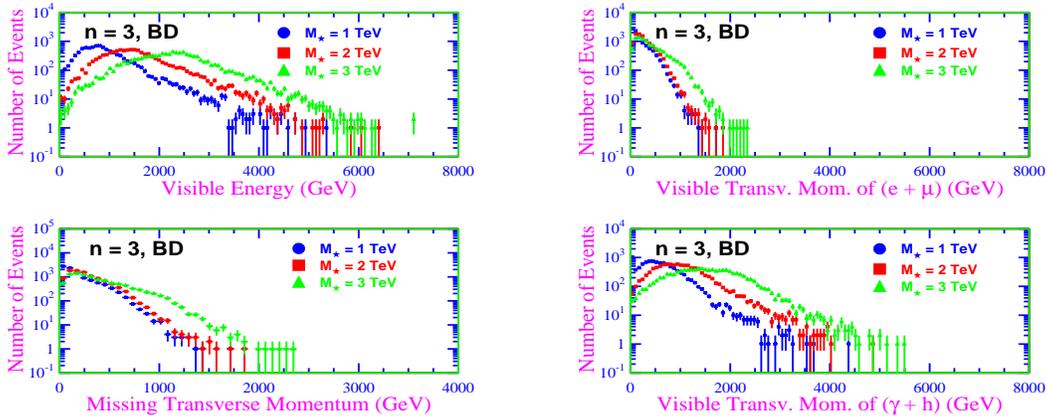}}
\vspace{-10truemm}
\caption[Visible energy, missing and visible transverse momentum for the BD model, $n=3$ and varying
$M_\star$.]{\label{figure:2} Visible energy, $\missPT$ and visible transverse momentum of leptons
and photons+jets (GeV) for the black disk model (BD) and fundamental Planck scale $M_\star=1,2,3$
TeV. The number of extra dimensions is $n=3$ and the final BH decay is in four hard quanta.}
\end{center}
\end{figure}

Figure \ref{figure:1} shows missing transverse momentum ($\missPT$) and visible transverse momentum
of leptons and hadrons for 10,000 events at the LHC with the following parameters (benchmark):
$$n=6\,,\qquad M_{min}=Q_{min}=M_\star\,,\qquad n_p=4\,,\qquad \alpha=0\,,$$
BD cross section and conservation of EM charge. The momentum transfer is chosen as the Schwarzschild
radius inverse. $P_T$ cuts of $5$ GeV on leptons ($e,\mu$) and  $15$ GeV on photons + hadrons
($\gamma,h$) have been imposed to remove the beams and inital-state radiation. (These choices of cuts
and momentum transfer apply to all simulations.) The plots show the total visible energy distribution,
$\missPT$ and the visible transverse momentum of leptons ($e,\mu$) and photons + jets ($\gamma,h$)
with varying fundamental scale $M_\star=1\dots 3$ TeV. Figure \ref{figure:2} shows the results for
three extra dimensions ($n=3$). The results in Fig.\ \ref{figure:1} and Fig.\ \ref{figure:2} are in
good agreement with simulations based on different BH generators \cite{Harris:2004xt}.

A handful of BH events shows a large amount of transverse momentum up to several TeV, depending
on the value of the fundamental scale and the number of extra dimensions. In the absence of a BH
remnant, this missing transverse momentum is due to the emission of gravitons and other invisible
quanta (e.g.\ neutrinos) in the various evolutionary phases of the BH (formation, Hawking
evaporation and final explosive phase). The bulk of BH events is characterized by light, low-entropy
BHs. Since the graviton and invisible channels accounts only for a small fraction of the total
multiplicity in the decay phase, only rare high-mass events show a large amount of missing
transverse momentum. A rough counting of degrees of freedom shows that the hadronic-to-leptonic
decay ratio of a BH event should be approximately 5:1. The prevalence of the hadronic channel on the
leptonic channel is evident from the right panels of Fig.\ \ref{figure:1} and Fig.\ \ref{figure:2}.
Figures \ref{figure:1} and \ref{figure:2} also show the effect of the fundamental scale on visible
energy and missing and visible transverse momentum. Increasing $M_\star$ leads to more massive BHs,
i.e., higher multiplicity and harder quanta in the Hawking phase. Therefore, higher values of
$M_\star$ tend to produce larger $\missPT$. Visible transverse momenta show a similar pattern.
Observation of events with high $\missPT$ would indicate high values of $M_\star$, independently of
the details of BH formation and the number of extra dimensions. If BHs are observed at the LHC,
$M_\star$ could be measured to a certain degree of precision. 

Missing and visible energy outputs depend on the initial BH mass, and thus from the number of extra
dimensions. Graviton emission in the Hawking phase also increases with $n$ \cite{Cardoso:2005vb},
leading to a decrease in visible energy for higher-dimensional BHs (compare the upper-left panels of
Fig.\ \ref{figure:1} and Fig.\ \ref{figure:2}.) However, the variation in $\missPT$ due to spacetime
dimensionality is much less significant than the change due to $M_\star$ because of the high degree
of sphericity of BH events (lower-left panels). Effects due to the dimensionality of spacetime are
more evident for massive BHs, whereas most of the BHs produced at the LHC are very light. Therefore,
it is unlikely that statistics alone will allow measurement of the number of extra dimensions. 

\begin{figure}[ht]
\begin{center}
\centerline{\includegraphics[height=0.4\textheight,width=0.92\textwidth]{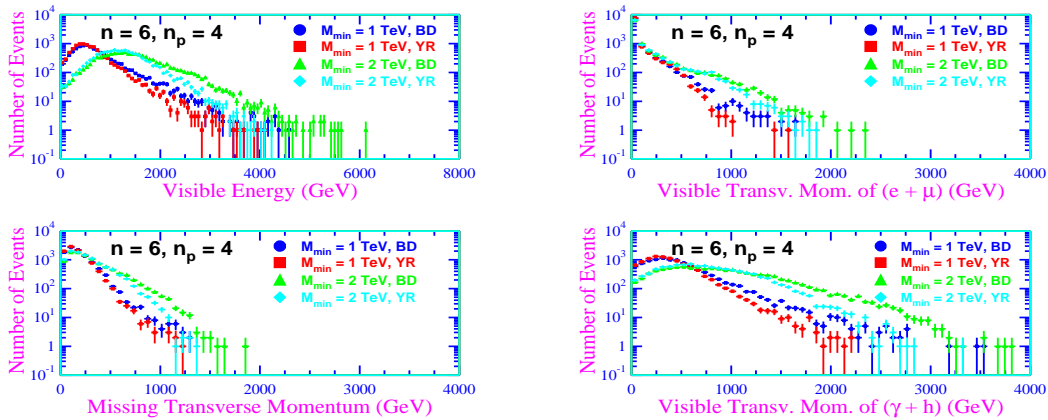}}
\vspace{-15truemm}
\caption[Visible energy, missing and visible transverse momentum for $M_\star=1$ TeV,
$n=6$ and varying $M_{min}$.]{\label{figure:3} 
Visible energy, $\missPT$ and visible transverse momentum of leptons and
photons+jets (GeV) for the black disk model (BD) and the Yoshino-Rychkov TS model
(YR) in a ten-dimensional spacetime ($n=6$) with fundamental Planck scale
$M_\star=1$ TeV. The minimum formation mass of the BH is $M_{min}=1$ TeV or
$M_{min}=2$ TeV. The final BH decay is in four hard quanta ($n_p=4$).}
\end{center}
\end{figure}

Figure \ref{figure:3} shows the effects of changes in the minimum mass cutoff. Simulations separate
quite easily different values of $M_{min}$. However, since $M_{min}$ is a lower bound on the BH
mass, increases in $M_{min}$ are akin to increases in $M_\star$ (compare the upper-left panels of
Fig.\ \ref{figure:1} and Fig.\ \ref{figure:3}). Changes in $M_{min}$ are also entangled with the
initial graviton emission, specially for massive events. In the BD model, larger values of $M_{min}$
(at fixed $M_\star$) lead to more massive BHs, and thus to higher visible transverse momenta. If the
initial gravitational emission is turned on, this increase may be balanced by a decrease due to
lower multiplicity (compare $M_{min}=1$ TeV for the BD model with $M_{min}=2$ TeV for the YR model).
A measure of $M_{min}$ might prove to be difficult at the LHC. 

\begin{figure}[ht]
\begin{center}
\centerline{\includegraphics[height=0.4\textheight,width=0.92\textwidth]{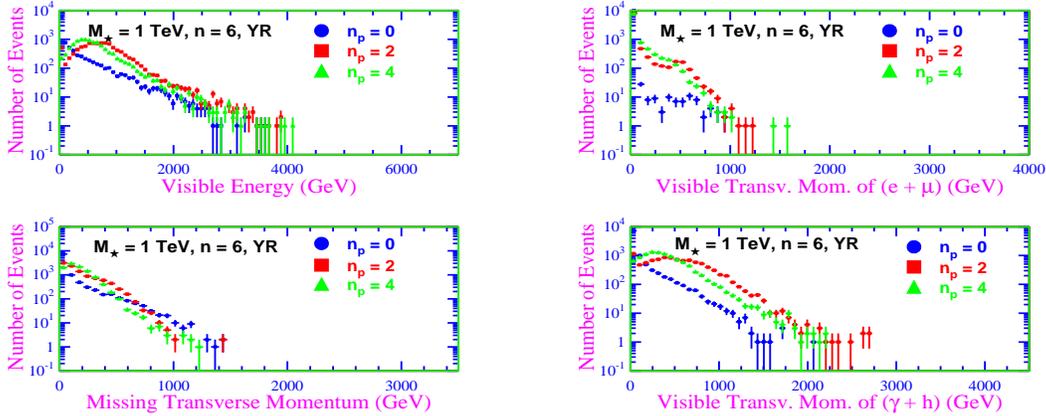}}
\vspace{-15truemm}
\caption[Visible energy, missing and visible transverse momentum for $M_\star=1$ TeV,
$n=6$ and varying final BH decay mode.]{\label{figure:4} Visible energy,
$\missPT$ and visible transverse momentum of leptons and photons+jets (GeV) for the
Yoshino-Rychkov TS model (YR) in a ten-dimensional spacetime ($n=6$) with fundamental Planck
scale $M_\star=1$ TeV and different final decay modes: neutral remnant ($n_p=0$),
two hard quanta ($n_p=2$) and four hard quanta ($n_p=4$).}
\end{center}
\end{figure}

Figure \ref{figure:4} displays the effects of the final explosive stage. Simulations show no
statistical difference between decay in $n_p=2$ and $n_p=4$ quanta. Since the degrees of freedom in
the final explosive phase are democratically chosen, a spectral analysis of the energy and the number
of emitted quanta is required to distinguish the two models. Detection of a BH remnant stands a better
chance because of larger $\missPT$ and smaller visible momentum due to the remnant undetectability.
(See also Refs.\ \cite{Koch:2005ks,Stocker:2006we}.) Note that a large fraction of events with remnant
produces very little visible output; most of the BHs are initially so light that the Hawking phase
does not take place. On the contrary, the energy carried by the decay products is much larger than the
invisible energy carried by the remnant for massive events.

\begin{figure}[ht]
\begin{center}
\centerline{\includegraphics[height=0.4\textheight,width=0.92\textwidth]{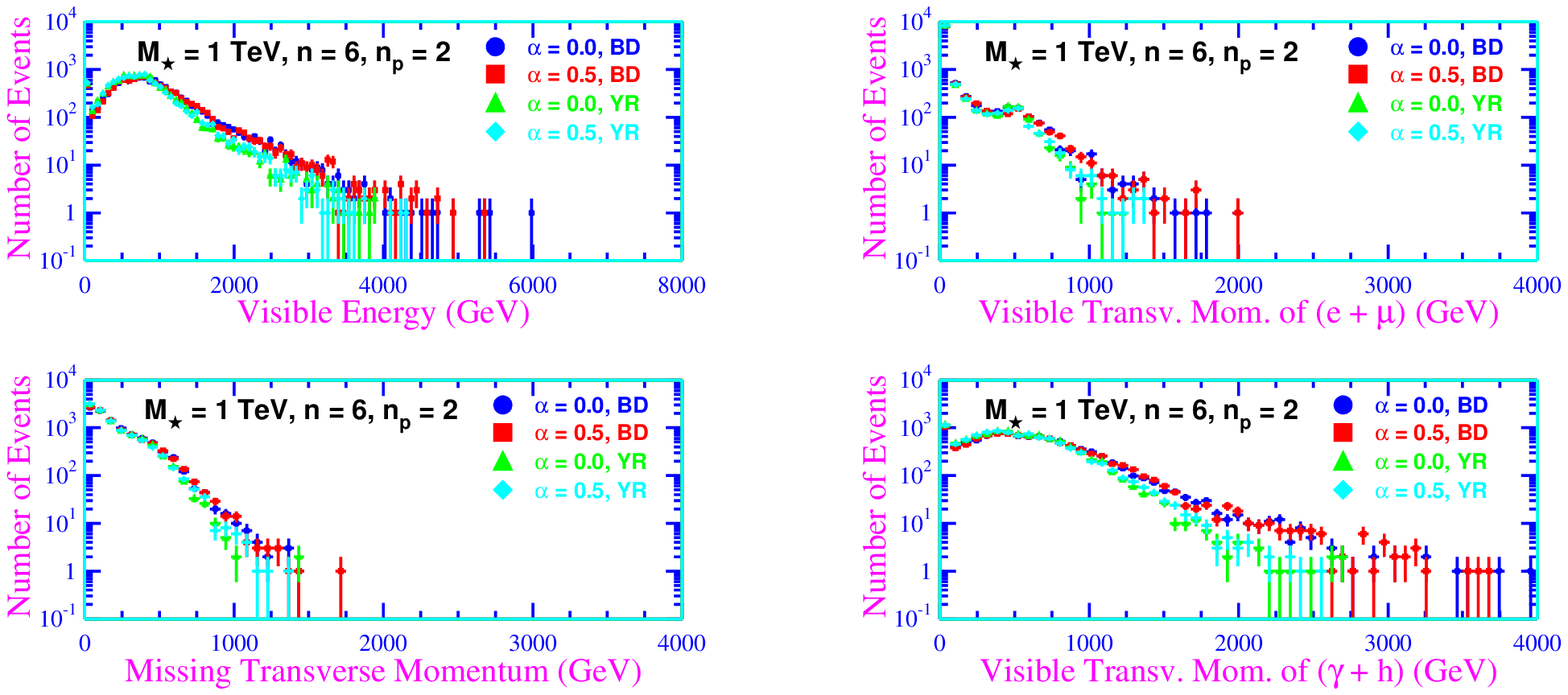}}
\vspace{-15truemm}
\caption[Visible energy, missing and visible transverse momentum for $M_\star=1$ TeV,
$n=6$ and nonzero minimum length.]{\label{figure:5}  Visible energy, $\missPT$ and
visible transverse momentum of leptons and photons+jets (GeV) for the black disk
model (BD) and the Yoshino-Rychkov TS (YR) model in a ten-dimensional spacetime
($n=6$) with fundamental Planck scale $M_\star=1$ TeV and zero ($\alpha=0$) or
$M_\star^{-1}$ ($\alpha=0.5$) minimum length. The final BH decay is in two hard
quanta ($n_p=2$).}
\end{center}
\end{figure}

Figure \ref{figure:5} compares BH events in a smooth spacetime ($\alpha=0$) and a spacetime with
minimum length equal to the fundamental Planck scale inverse ($\alpha=0.5$). The simulations show no
significant statistical differences between the two cases. The effects of a small distance cutoff
becomes only relevant when the minimum scale is very close to the threshold of complete suppression of
BH production. In this case, the minimum allowed mass Eq.\ (\ref{minmass}) is so large that BHs cannot
form at the LHC CM energy. Therefore, observation of minimum length effects at the LHC requires a
certain degree of fine tuning. It is unlikely that any information on quantum effects at the Planck
scale can be extracted from LHC data.
\subsection{Event shape} 
\begin{figure}[ht]
\begin{center}
\centerline{\includegraphics[height=0.55\textheight,width=0.55\textwidth]{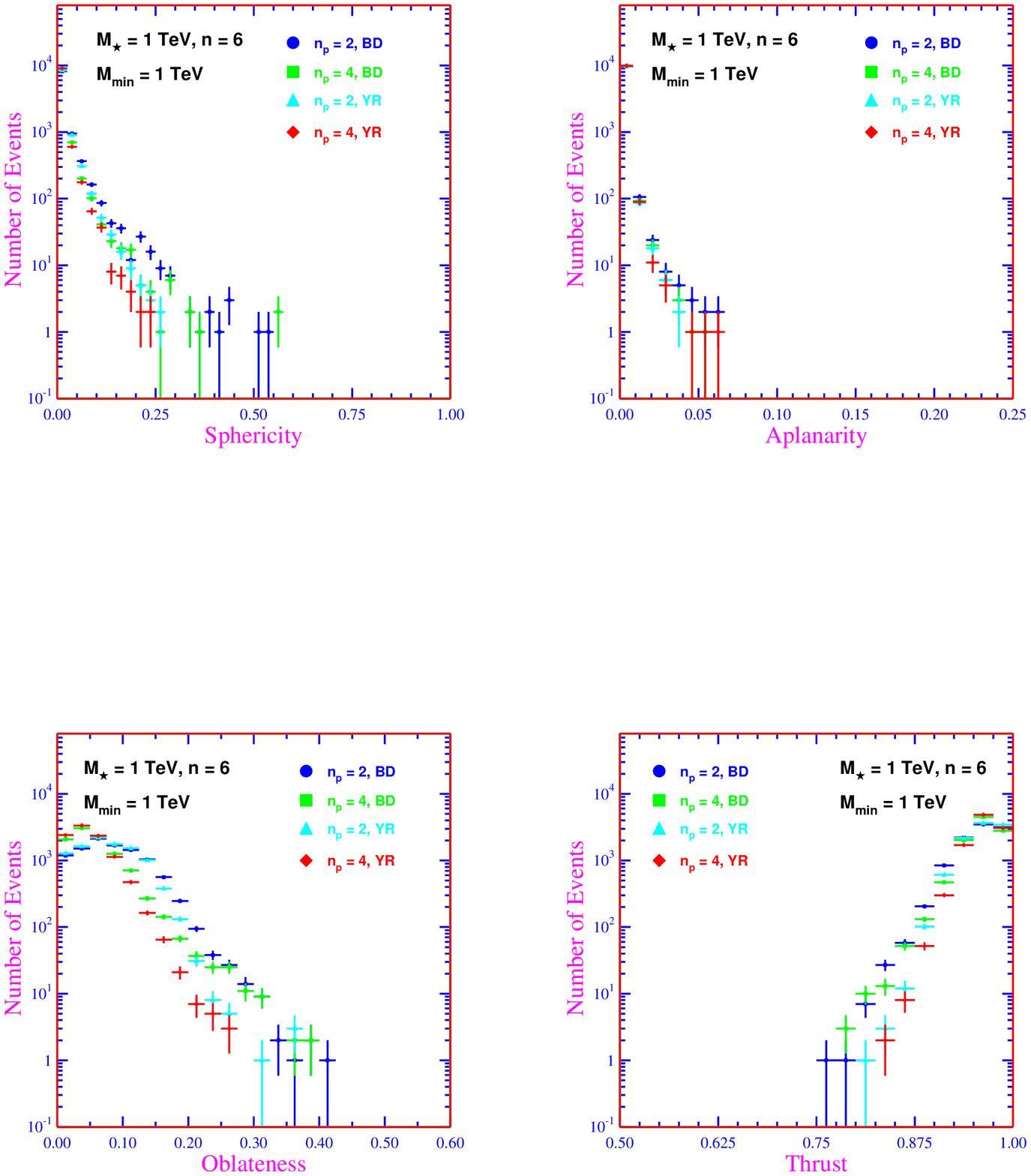}}
\caption[Sphericity, aplanarity, oblateness and thrust for
$M_\star=1$ TeV, $n=6$ and varying final BH decay mode.]{\label{figure:6} 
Sphericity, aplanarity, oblateness and thrust for the black
disk model (BD) and the Yoshino-Rychkov TS model (YR) in a ten-dimensional
spacetime ($n=6$). The final black hole decay is in two hard
quanta ($n_p=2$) or four hard quanta ($n_p=4$).}
\end{center}
\end{figure}

BH events are expected to be highly spherical because of the spherical nature of Hawking
evaporation. The event shape can be quantified by means of the sphericity $S$ and aplanarity $A$
\cite{Bjorken:1969wi}, thrust and oblateness $T$ \cite{Brandt:1964sa}, and Fox-Wolfram moment
$R_1\dots R_4$ variables \cite{Fox:1978vu}. Fig.\ \ref{figure:6} shows sphericity, aplanarity,
oblateness and thrust for a ten-dimensional model with fundamental Planck scale equal to 1 TeV,
$M_{min}=Q_{min}=M_\star$, no minimum length, different formation and final decay models. (Rare)
massive BH events are characterized by very high sphericity and isotropy. A similar conclusion is
reached by examining the second Fox-Wolfram moment (see first panel of Fig.\ \ref{figure:7}).
Increasing $M_{min}$ makes the events even more spherical because of the higher multiplicity in the
decay phase.

Comparison between formation models at fixed $n_p$ shows that more spherical events are obtained if
the graviton loss is neglected; BHs are more massive and emit more quanta in the Hawking phase. The
higher sphericity of BD events is evident from the central-right part of the plots, where Hawking emission
dominates the emission in the final explosive phase. This makes the statistical difference between
the formation models more clear. Comparison between $n_p=2$ and $n_p=4$ at given formation model
shows that the former are less spherical than the latter. This effect is better
displayed in the region of the plots corresponding to light BHs, where emission in the final phase
dominates over Hawking emission. However, it should be stressed that the distinction between $n_p=2$
and $n_p=4$ at the LHC might be difficult due to the presence of non-BH background (e.g.\ $q\bar q$
events). Discrimination between alternative models of BH formation should be possible by selecting
massive spectacular events with high sphericity. 
\subsection{Jet parameters} 
\begin{figure}[ht]
\begin{center}
\centerline{\includegraphics[height=0.55\textheight,width=0.55\textwidth]{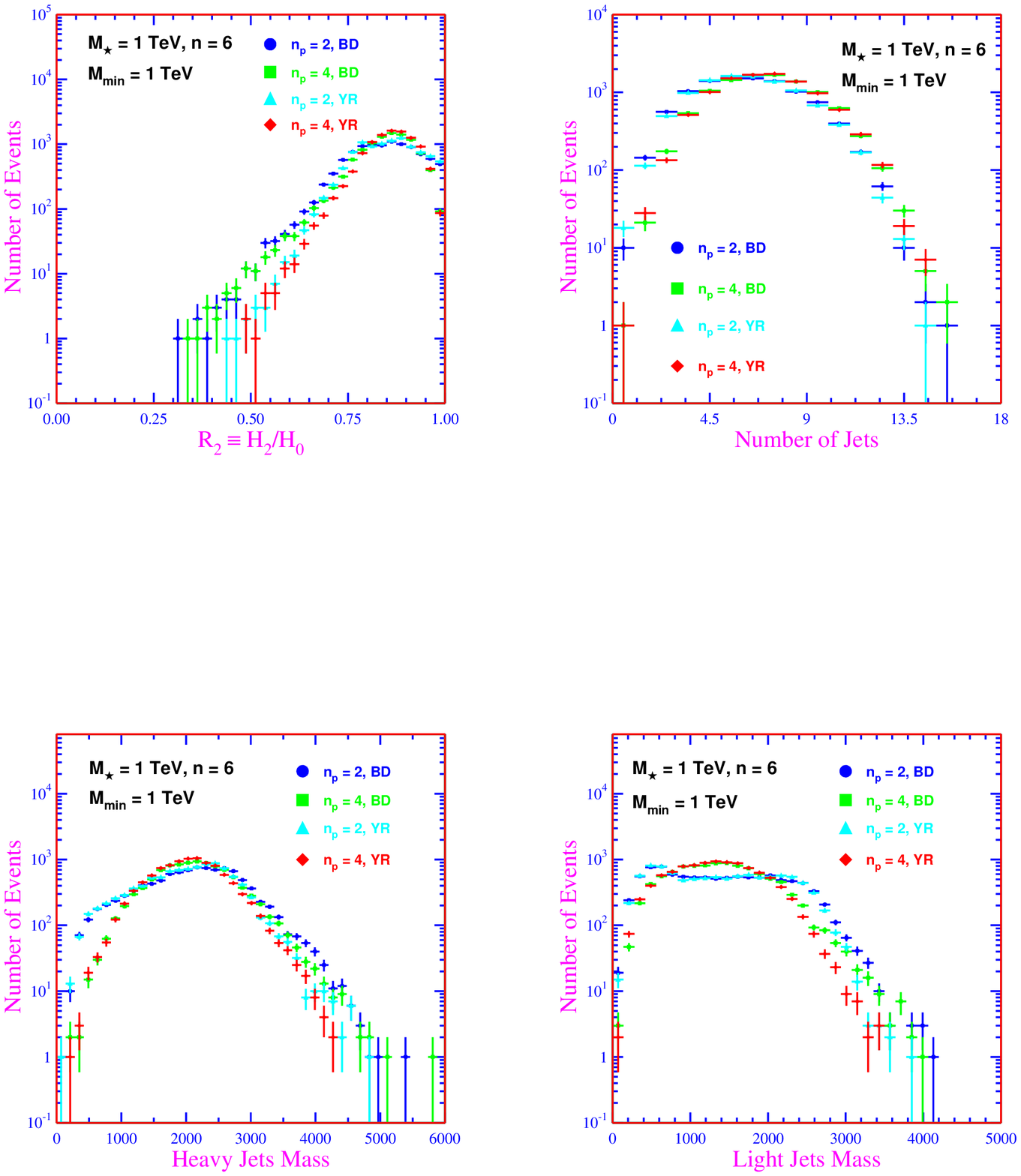}}
\caption[Fox-Wolfram moment $R_2$, number of jets, heavy and light jet mass for
$M_\star=1$ TeV, $n=6$ and varying final BH decay mode.]{\label{figure:7}
Fox-Wolfram moment $R_2$, number of jets, heavy and light jet mass for the black
disk model (BD) and the Yoshino-Rychkov TS model (YR) in a ten-dimensional
spacetime ($n=6$). The final black hole decay is in two hard quanta ($n_p=2$) or
four hard quanta ($n_p=4$).}
\end{center}
\end{figure}

The upper-right and the lower panels of Fig.\ \ref{figure:7} show the number of jets and the heavy
and light jet mass \cite{Sjostrand:2006za} for the choice of parameters discussed above,
respectively. These plots include initial- and final-state radiation jets in addition to the jets
originated in the BH decay phase. As is expected, the BD model produces on average more jets than
the model with graviton loss at formation (upper-right panel of Fig.\ \ref{figure:7}). This is also
evident from the right portions of the jet mass distributions, where the BD model is characterized
by more massive jets than the YR model at fixed $n_p$. Therefore, measurement of high jet mass
allows determination of the BH formation model independently of the shape variables. The left
portions of the jet mass distributions are sensitive to the final BH decay. Final decay in $n_p=2$
jets produces more heavy jets than final decay in $n_p=4$ jets. Therefore, the measurement of low
jet mass may give important information on the physics of the final BH phase.

\section{Conclusions and further developments}
The study of BH production at the TeV scale is now a few years old and entering the mature stage.
With the LHC scheduled to begin operations soon, accurate simulations of BH events are a pressing
need. These simulations should check the stability of the overall picture of BH production against
improvements in the theory and give independent confirmation of previous results. In this paper we
have investigated the signatures of BH events at the LHC with the MC generator CATFISH. CATFISH
implements several features of BH production at the TeV scale which were not included in previous
generators \cite{Gingrich:2006gs}. Our analysis has shown that the main signatures of BH production
at the LHC (missing transverse momentum, high sphericity, high jet multiplicity) do not depend
significantly on the fine details of BH formation and evolution. Measurement of the fundamental
Planck scale and detection of a BH remnant could possibly be extracted from LHC data. On the other
hand, discerning different models of BH formation and evolution at the LHC might prove difficult on
a purely statistical basis  

Several other interesting signatures of BH formation in particle colliders have been proposed in the
literature (see, e.g., Refs.\ \cite{Stocker:2006we, Harris:2004xt, Tanaka:2004xb, Lonnblad:2005ah,
Nayak:2006vf}). In particular, suppression of high-energy back-to-back-correlated di-jets with
energy above the fundamental scale and di-lepton production with large transverse momentum are
expected to be two of the most interesting signatures of BH production at the LHC. Investigation of
these signatures with CATFISH is in progress. Detector response and event
reconstruction are also fundamental issues to be addressed in a complete analysis of BH events at
the LHC. Further work along these lines is currently being pursued. 
\section*{Acknowledgments}
The authors would like to thank Vitor Cardoso, David Cline, Greg Landsberg, Alexander Melnitchouk,
Robert Palmer, Hans Wenzel and Graham Wilson for discussions and many useful suggestions. This work
was supported in part by the U.S. Department of Energy contract  DE-FG05-91ER40622.

\end{document}